\documentclass[fleqn,twoside]{article}
\usepackage{espcrc2}

\usepackage{amssymb}
\usepackage{epsfig}

\usepackage[figuresright]{rotating}

\newcommand{\fpi}{F_{\pi}}
\newcommand{\mpi}{M_{\pi}}
\newcommand{\eq}[1]{eq.\,(\ref{#1})}

\title{\vspace{-8mm}
{\normalsize DESY 02-131\hfill{\tt hep-lat/0209110}}\\[-2mm]
{\normalsize September 2002}\\[5mm]
Finite size effects on $M_\pi$ in QCD from Chiral Perturbation Theory%
\thanks{presented by S.\ D\"urr at Lattice 02, Boston MA.}%
\thanks{We thank the INT in Seattle for kind hospitality, since our work was
initiated during the program INT-01-03.}%
\thanks{Two of us (G.C.\ and R.S.) thank the EURIDICE network for support
under contract HPRN-CT-2002-00311.}}

\author{
G. Colangelo\address{Institute for theoretical Physics, University of Bern,
3012 Bern, Switzerland},
S. D\"urr\address[DESY]{DESY, Platanenallee 6, 15738 Zeuthen, Germany},
R.\ Sommer\addressmark[DESY]}

\begin{document}

\begin{abstract}
We present a determination of the shift $M_\pi(L)-M_\pi$ due to the finite
spatial box size $L$ by means of $N_\mathrm{f}\!=\!2$ Chiral Perturbation
Theory and L\"uscher's formula. The range of applicability of the chiral
prediction is discussed.
\end{abstract}

\maketitle


\section{INTRODUCTION}

Finite size effects are important systematic effects in the Monte Carlo
treatment of lattice field theories~\cite{Fukugita:1992hr}.
They become particularly large when the spectrum contains light particles as it
happens for QCD with light quark masses.
Fortunately, chiral symmetry allows to investigate this region analytically by
chiral perturbation theory (CHPT)~\cite{Gasser:1983yg,GaLeFSE1,betterways}.
In fact, through the detour of the chiral Lagrangian, information about
infinite volume QCD may be obtained from finite volume simulations with box
size $L$~\cite{eps_exp_use}.
An important restriction is that
\begin{equation}
 \fpi L \gg 1\,, \quad \mpi/(4\pi \fpi) \ll 1
 \label{e_largel}
\end{equation}
has to be satisfied, since the chiral theory is a low-energy (long distance)
expansion.
In addition, 
one has to distinguish whether $L$ is large compared to the Compton wave
length of the pion or not~\cite{betterways}.
We shall here restrict ourselves to the former case but investigate what
\eq{e_largel} means quantitatively, by considering more than the leading order
(LO) in the chiral expansion.

More specifically, we study the pion mass, $\mpi(L)$, defined as an eigenvalue
of the QCD Hamiltonian in an $L\times L\times L$ box (periodic boundary
conditions), as it is extracted on a Euclidean lattice for sufficiently large
time.
The goal is to calculate the shift $M_\pi(L)-M_\pi$ with
$M_\pi=M_\pi(L\!=\!\infty)$.


\section{L\"USCHER FORMULA}

\begin{figure}[!b]
\vspace{-4mm}
\begin{center}
\unitlength 0.6cm
\begin{picture}(12,6)(0,0)
\linethickness{0.2mm}
\put(0,3){\line(1,0){6.5}}
\put(7,3){\line(1,0){005}}
\put(6,0){\line(0,1){6}}
\put(1,5){\oval(1.5,1.5)}
\put(0.85,4.85){$\nu$}
\linethickness{1.6mm}
\put(00,3){\line(1,0){2}}
\put(10,3){\line(1,0){2}}
\put(1.5,2){\large $-M_\pi$}
\put(9.5,2){\large $ M_\pi$}
\linethickness{0.6mm}
\put(6,0.5){\line(0,1){5}}
\put(5.79,5.5){$\blacktriangle$}
\put(6.5,1.4){\begin{turn}{90}{\normalsize\bf Integration}\end{turn}}
\end{picture}
\end{center}
\vspace{-12mm}
\caption{Integration contour in the complex $\nu$ plane where $\nu$ is the
crossing variable in the (Minkowski space) forward scattering amplitude.}
\end{figure}
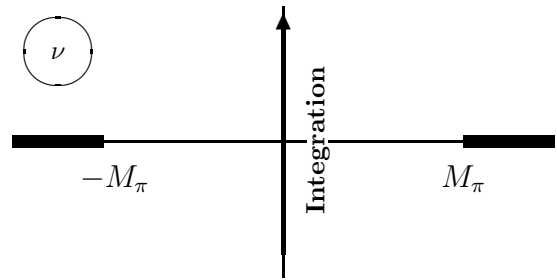

The formula that we will use \cite{Luscher:1985dn},
\begin{eqnarray}
M_\pi(L)\!\!&\!\!-\!\!&\!\!M_\pi=-{3\over16\pi^2M_\pi L}\times
\label{luscher}
\\
&{\Big(}\!\!&\!\!\!
\int_{-\infty}^\infty\!\!dy\:e^{-\sqrt{M_\pi^2+y^2}L}F({\rm i}y)+
O(e^{-\overline{M}L})
\Big),
\nonumber
\end{eqnarray}
relates the finite volume mass shift to the $\pi$-$\pi$ forward scattering
amplitude $F$ in infinite volume.
The subleading term involves $\overline{M}\!>\!\sqrt{3/2}\:M_\pi$.
An additive piece referring to the 3-particle vertex present in the original
formula \cite{Luscher:1985dn} does not contribute,
because of parity conservation in QCD.
Since \eq{luscher} builds on the unitarity of the theory, it holds only in full
QCD, while a generalization to the quenched approximation seems impossible.

A particular virtue of the L\"uscher formula is an economic one: inserting on
the r.h.s.\ the {\em one-loop\/} expression in CHPT for $F$ in {\em infinite
volume\/} and integrating along the contour depicted in Fig.\ 1, one gets the
asymptotic part of a formula which, otherwise, would have required a
{\em two-loop\/} computation in a {\em finite volume\/}.


\section{INPUT FROM CHIRAL PT}

The simple $a,b \rightarrow c,d$ forward kinematics
\begin{equation}
\begin{array}{cclcl}
s(\nu)&=&(p^{(a)}+p^{(b)})^2&=&2M_\pi^2+2M_\pi\nu\\
t(\nu)&=&(p^{(a)}-p^{(c)})^2&=&0\\
u(\nu)&=&4M_\pi^2-s&=&2M_\pi^2-2M_\pi\nu
\end{array}
\end{equation}
with the crossing variable
\begin{equation}
\;\; \nu \equiv {p^{(a)}p^{(b)}\over M_\pi}={s\over2M_\pi}-M_\pi
\label{cross}
\end{equation}
means that the forward scattering amplitude, after summing over isospin, reads
\begin{equation}
\begin{array}{rl}
F(\nu)=\!\!&\!\!A(s(\nu),0,u(\nu))+A(u(\nu),0,s(\nu))+
\\
{}&\!\!3A(0,s(\nu),u(\nu))\:,
\end{array}
\label{forw}
\end{equation}
where the isospin-invariant $\pi$-$\pi$ scattering amplitude $A(s,t,u)$ as
defined in \cite{Gasser:1983yg} enters.

We now use the expression for $A$ at either leading or next-to-leading order
\cite{Gasser:1983yg}, construct $F(\nu)$ via \eq{forw}, and insert the result
into \eq{luscher}.
At LO, this procedure has already been used in \cite{Fukugita:1992hr}.


\section{MASS SHIFTS AT LO/NLO}

The LO chiral expression for the amplitude \cite{Gasser:1983yg}
\begin{equation}
A(s,t,u)\vert_{\rm LO}=(s-M_\pi^2)/F_\pi
\label{alo}
\end{equation}
depends only on $s$, and henceforth the forward amplitude \eq{forw} is a
constant and the relative shift
\begin{eqnarray}
\label{e_LOshift}
{M_\pi(L)-M_\pi\over M_\pi}\Big\vert_{\rm LO}
&=&
{3\over8\pi^2}\,{M_\pi^2\over F_\pi^2}\,
{K_1(M_\pi L)\over M_\pi L}\\
{}&\sim&
{3\over4(2\pi)^{3/2}}\,{M_\pi^2\over F_\pi^2}\,
{e^{-M_\pi L}\over (M_\pi L)^{3/2}}
\nonumber
\end{eqnarray}
follows in closed form.
As noticed by Gasser and Leutwyler, this expression agrees with their one-loop
result \cite{GaLeFSE1} for $\mpi(L)$, when the non-leading terms
$O(e^{-\overline{M}L})$ are dropped in the latter (here,
$\overline{M}\!=\!\sqrt{2}\mpi$).
In addition, at that order of CHPT, one can check explicitly that the
non-leading terms get very small around $\mpi L\!\sim\!4$.

We now investigate, whether the subleading chiral corrections in $A$ lead to
a sizeable modification of the LO mass shift \eq{e_LOshift}.
At NLO the chiral expression for the scattering amplitude is more involved
than \eq{alo} \cite{Gasser:1983yg}, and we evaluated the integral in
\eq{luscher} numerically.
The low-energy constants which are needed at NLO are taken from
\cite{Colangelo:2001df}, and -- for the time being -- the influence of their
uncertainties is not investigated.


It is instructive to first look at the integrand
$I(y)\!=\!\exp(-\sqrt{M_\pi^2+y^2}\,L)\,F(\mathrm{i}y)$ in \eq{luscher}.
In Fig.\ 2 we plot it, with $F$ constructed via LO and NLO CHPT, respectively.
Since the calculation is based on CHPT and the integration variable $y$ has
the dimension of a mass, either the integrand tends to zero sufficiently fast
beyond $O(100)\,$MeV or the calculation cannot be trusted.

\begin{figure}[!b]
\vspace{-4mm}
\begin{center}
 \epsfig{file=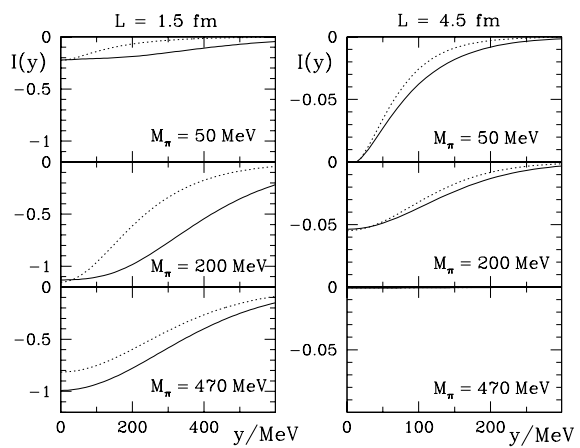,width=7.5cm}
\end{center}
\vspace{-10mm}
\caption{The integrand $I(y)$ at LO (dotted) and NLO (full line).}
\end{figure}

\begin{figure*}[t!]
\begin{center}
 \epsfig{file=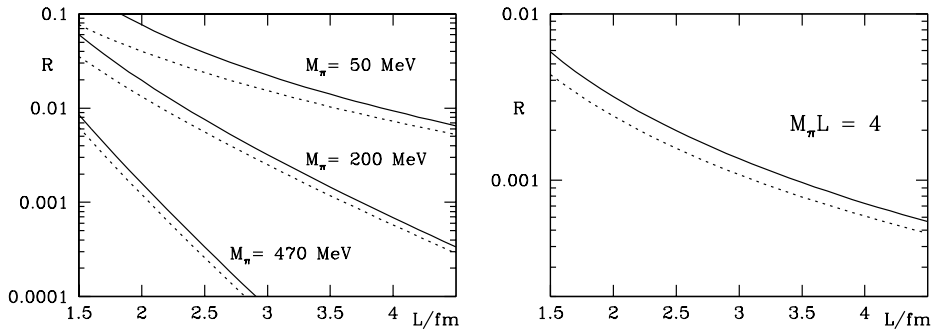,width=12.3cm}
\end{center}
\vspace{-12mm}
\caption{$R\!=\!(M_\pi(L)\!-\!M_\pi)/M_\pi$ (with
$M_\pi\!\equiv\!M_\pi(L\!=\!\infty))$ from LO (dotted)
and NLO (full line) CHPT.}
\end{figure*}

The two integrands are close to each other when the pion is light and the box
size sufficiently large.
However, in the whole range of $L$ and $\mpi$ studied in the graphs, the
relative difference is substantial.
The reason is that at NLO $F(\mathrm{i}y)$ contains contributions which grow
quite strongly with $y$.
To suppress these, a large $L$ in the kinematic factor
$\exp(-\sqrt{\mpi^2+y^2}L)$ is needed.
Indeed, one roughly needs $\fpi L\!\geq\!3$ and $\mpi/\fpi\!\leq\!4$, if one
requires $I_\mathrm{NLO}$ to differ only by a few percent from $I_\mathrm{LO}$.

Note that in the graphs in Fig.\ 2 also values of $\mpi L$ appear which are
too small for the asymptotic formula \eq{luscher} to apply.
We do this in order to check which value of $L$ is needed for a precise
prediction for the {\em integrand\/} from CHPT.
In addition, it is an attempt to shed some light on the general question,
what the condition $\fpi L\!\gg\!1$ in \eq{e_largel} means numerically.
This question is relevant also for applications of CHPT in the regime
$\mpi L\!\ll\!1$, where \eq{luscher} has no basis but where other expansions
\cite{betterways,eps_exp_use} are applicable.


The relative mass shift $R=\!(M_\pi(L)\!-\!M_\pi)/M_\pi$ is plotted in Fig.\ 3.
One observes that for $\mpi L\!=\!4$ and $L\!=\!1.5\,$fm the predicted mass
shift is significantly below 1\% and decreases further when $L$ is increased.
Even though the difference between LO and NLO is not so small, we expect this
to hold beyond NLO, on the basis of our preliminary numerical study of NNLO
effects.

Larger mass shifts are predicted for a pion mass of e.g.\ $200\,$MeV, and box
sizes below $2.5\,$fm.
In this region, one would then like to correct lattice MC results for such
effects, but it seems that the mass shifts of NLO CHPT may only be trusted
within around 20\% (of the very shift).

We did not study $L\!<\!1.5\,$fm, since in this region, the ``distortion of the
pion wave function'' due to finite $L$ may be large~\cite{Fukugita:1992hr},
and we expect that higher order CHPT contributions will at most describe the
onset of such effects.

\bigskip

In summary, it appears that CHPT may be applied to find out the region of
parameters where the finite size effects are small and negligible compared to
other errors of lattice QCD.
We do, however, find that in the region where {\em a correction for finite size
effects is necessary\/}, a {\em LO CHPT result is not sufficient\/} to apply
this correction with confidence.
In order to really assess the precision of the NLO prediction, it will be
interesting to include yet one more order in the scattering amplitude
\cite{Colangelo:2001df}, but then the uncertainties in the needed low-energy
constants have to be investigated in detail.
Such work is in progress.


\end{document}